\documentclass[aps,amssymb,prd,amssymb]{revtex4}
\usepackage{graphicx}
\newcommand{\be}{\begin{equation}}
\newcommand{\ee}{\end{equation}}
\newcommand{\ba}{\begin{eqnarray}}
\newcommand{\ea}{\end{eqnarray}}

\begin{document}
\title{A Singular Conformal Spacetime}

\author{R. Aldrovandi}
%\email{ra@ift.unesp.br}

\author{J. P. Beltr\'an Almeida}
%\email{jalmeida@ift.unesp.br}

\author{J. G. Pereira}
%\email{jpereira@ift.unesp.br}

\affiliation{Instituto de F\'{\i}sica Te\'orica,
Universidade Estadual Paulista \\
Rua Pamplona 145, 01405-900 S\~ao Paulo SP, Brazil}
%\date{\today}

\begin{abstract}
The infinite cosmological ``constant'' limit of the de Sitter solutions to Einstein's equation is
studied. The corresponding spacetime is a singular, four-dimensional cone-space, transitive under {\it
proper} conformal transformations, which constitutes a new example of maximally-symmetric spacetime.
Grounded on its geometric and thermodynamic properties, some speculations are made in connection with
the primordial universe.
\end{abstract}

%\pacs{04.20.Cv; 98.80.Es; 98.80.Bp}

\maketitle

%%%%%%%%%%%%%%%%%%%%%%
\section{Introduction}

The de Sitter and anti-de Sitter spacetimes are the only possible uniformly {\em curved}
four-dimensional metric spacetimes \cite{weinberg}. They are maximally--sym\-me\-tric, in the sense
that they can lodge the maximum number of Killing vectors. These spacetimes are related respectively to
a positive and to a negative cosmological constant $\Lambda$, and their groups of motions are
respectively the de Sitter and anti-de Sitter groups. In the limit of a vanishing $\Lambda$, both de
Sitter and anti-de Sitter groups are contracted \cite{inonu1} to the Poincar\'e group.  Concomitantly,
both de Sitter and anti-de Sitter spacetimes are reduced to the Minkowski space, a {\em flat}
maximally-symmetric spacetime. In the contraction procedure, therefore, the initial and final states
are maximally-sym\-me\-tric spacetimes.

A less studied contraction is the opposite limit of an infinite cosmological constant \cite{mwa}.  As in
the previous case, the degree of symmetry is preserved in this contraction process, resulting in a new
maximally-symmetric spacetime, with a well defined group of motions.  It is a four-dimensional cone
spacetime, here denoted $N$, singular at the cone vertex. Similarly to the Poincar\'e group, which is
the group of motions of the Minkowski spacetime, the group ruling the kinematics of the cone spacetime
$N$ is the so called {\it conformal} Poincar\'e group, the semi-direct product of Lorentz and the {\it
proper} conformal groups \cite{ap1}.  Despite presenting a well defined algebra, however, the spacetime
metric becomes unavoidably singular, precluding the existence of any notion of space distance and time
interval.  Nevertheless, a conformal invariant metric can be defined, which allows the definition of
{\it conformal} notions of space distance and time interval. Except at the singular cone vertex,
therefore, this new maximally-symmetric spacetime can be considered as conformally smooth, and infinite
from the point of view of the conformal notions of space and time translations. Differently from the
Minkowski spacetime, which is transitive under ordinary spacetime translations, this new
maximally-symmetric spacetime is found to be transitive under spacetime proper conformal
transformations.

In this work, we will explore the geometric properties of the singular, conformally infinite, cone
spacetime $N$.  Relying upon these features, as well as on its thermodynamic properties, a speculation
about possible cosmological applications of these ideas will then be made. We begin by reviewing, in the
next section, the fundamentals of de Sitter spaces and groups.

%%%%%%%%%%%%%%%%%%%%%%%%%%%%%%%%%%%%%%%%%
\section{The de Sitter Spaces and Groups}

\subsection{Definitions and Notations} 

There are two different de Sitter spacetimes \cite{ellis}, one with negative, and one with
positive scalar curvature. They can be defined as hypersurfaces in the pseudo-Euclidean spaces ${\bf
E}^{4,1}$ and ${\bf E}^{3,2}$, inclusions whose points in Cartesian coordinates $(\xi^A) = (\xi^0,
\xi^1, \xi^2, \xi^3$, $\xi^{4})$ satisfy, respectively,
\[
\eta_{AB} \xi^A \xi^B \equiv (\xi^0)^2 - (\xi^1)^2 - (\xi^2)^2 - (\xi^3)^2 - (\xi^{4})^2 =
- L^2
\]
and
\[
\eta_{AB} \xi^A \xi^B \equiv (\xi^0)^2 - (\xi^1)^2 - (\xi^2)^2 - (\xi^3)^2 + (\xi^{4})^2 =
L^2,
\]
with $L$ the so called de Sitter length parameter. We use $\eta_{a b}$ ($a, b = 0,1,2,3$) for the
Lorentz metric $\eta = $ diag $(1$, $-1$, $-1$, $-1)$, and the notation ${\sf s} = \eta_{44}$ to put
both the above conditions together as
\be
\eta_{a b} \, \xi^{a} \xi^{b} + {\sf s} \left(\xi^4\right)^2 = {\sf s} \, L^2.
\label{dspace2}
\ee
For ${\sf s} = - 1$, we have the de Sitter space $dS(4,1)$, whose metric is derived from the
pseudo-Euclidean metric $\eta_{AB}$ = $(+1,-1,-1,-1,-1)$. It has the group $SO(4,1)$ as group of
motions. For ${\sf s} = + 1$, we have the so called anti-de Sitter space, denoted by $dS(3,2)$. It
comes from $\eta_{AB}$ = $(+1,-1,-1,-1,+1)$, and has $SO(3,2)$ as its group of motions. Both spaces
are homogeneous:
\[
dS(4,1) = SO(4,1)/ SO(3,1) \quad {\rm and} \quad dS(3,2) = SO(3,2)/ SO(3,1).
\]
In addition, each group manifold is a bundle with the corresponding de Sitter or anti-de Sitter space
as base space, and the Lorentz group ${\mathcal L}=SO(3,1)$ as fiber \cite{kono}.

\subsection{Stereographic Coordinates}

The four-dimensional stereographic coordinates are obtained through a ste\-re\-o\-graphic projection
from the de Sitter hypersurfaces into a Minkowski spacetime. It is defined by \cite{gursey}
\be
\xi^{a} = h^{a}{}_{\mu} \, x^\mu \equiv \Omega(x) \, \delta^{a}{}_{\mu} \, x^\mu
= \Omega(x) \, x^a
\label{xix}
\ee
and
\be
\xi^4 = - L \, \Omega(x) \left(1 - {\sf s} \,
\frac{\sigma^2}{4 L^2} \right),
\label{xi4}
\ee 
where
\be
\Omega(x) = \frac{1}{1 + {\sf s} \, {\sigma^2}/{4 L^2}},
\label{n}
\ee
with $\sigma^2 = \eta_{a b} \, x^a x^b = \eta_{\mu \nu} \, x^\mu x^\nu$ the Lorentz invariant interval.
In these expressions we have used the relations $x^a = \delta^{a}{}_{\mu} \, x^\mu$ and $\eta_{\mu
\nu} = \delta^{a}{}_{\mu} \, \delta^{b}{}_{\nu} \, \eta_{a b}$. The $h^{a}{}_{\mu}$ introduced in
(\ref{xix}), on the other hand, are components of a nontrivial tetrad field, actually of the 1-form
basis members
\[
h^{a} = h^{a}{}_{\mu} dx^\mu = \Omega(x) \, \delta^{a}{}_{\mu} dx^\mu = \Omega(x) \, dx^a.
\]
In terms of the stereographic coordinates, therefore, the de Sitter metric is
\be
g_{\mu \nu} = h^{a}{}_{\mu} \, h^{b}{}_{\nu} \, \eta_{a b} \equiv \Omega^2(x) \, \eta_{\mu \nu}.
\label{44}
\ee
The de Sitter spaces are thus conformally flat, with the conformal factor given by $\Omega^2(x)$. Notice
that we are carefully using the Latin alphabet for the algebra (nonholonomous) indices, and the Greek
alphabet for the (holonomous) homogeneous space fields and cofields. As usual with changes from flat
tangent-space to spacetime, indices of the two kinds are interchanged with the help of a tetrad field.

The de Sitter metric (\ref{44}) is a solution of the sourceless Einstein's equation
\be
R_{\mu \nu} - \frac{1}{2} g_{\mu \nu} \, R - \Lambda \, g_{\mu \nu} = 0,
\ee
where the cosmological term $\Lambda$ and the length parameter $L$ are related by
\be
\Lambda = - \frac{3 {\sf s}}{L^2}.
\label{lambdaR}
\ee
The energy density associated with the cosmological term is, therefore,
\be
\varepsilon_\Lambda = \frac{c^4 \Lambda}{8 \pi G} = -\, {\sf s} \, \frac{3 c^4}{8 \pi G L^2}.
\label{darked}
\ee
The Christoffel connection of the metric (\ref{44}) is \cite{livro}
\be
\Gamma^{\lambda}{}_{\mu \nu} = \left[ \delta^{\lambda}{}_{\mu}
\delta^{\sigma}{}_{\nu} + \delta^{\lambda}{}_{\nu}
\delta^{\sigma}{}_{\mu} - \eta_{\mu \nu} \eta^{\lambda \sigma} \right]
\partial_\sigma [\ln \Omega(x)],
\label{46}
\ee
with the Riemann tensor components given by
\be
R^{\mu}{}_{\nu \rho \sigma} = -\, \frac{\Lambda}{3} \,
\left[\delta^{\mu}{}_{\rho} g_{\nu \sigma} - \delta^{\mu}{}_{\sigma} g_{\nu
\rho} \right].
\label{47}
\ee
The Ricci and the scalar curvature tensors are, consequently,
\be
R_{\mu \nu} = -\, \Lambda \, g_{\mu \nu} \quad
{\rm and} \quad R = -\, 4 \Lambda.
\label{49}
\ee 
Observe that, according to our convention, the de Sitter (anti-de Sitter) space has negative (positive)
scalar curvature. Of course, both of them have negative Gaussian curvature. 

\subsection{Kinematic Groups: Transitivity}

The kinematic group of any spacetime will always have a subgroup accounting for both the isotropy of
space (rotation group) and the equivalence of inertial frames (boosts). The remaining transformations,
generically called {\it translations}, can be either commutative or not, and are responsible for the
homogeneity of space and time. This holds of course for usual Galilean and other conceivable
non-relativistic kinematics \cite{levy}, but also for special-relativistic kinematics. The best known
relativistic example is the Poin\-ca\-r\'e group ${\mathcal P}$, naturally associated with the
Minkowski spacetime $M$ as its group of motions. It contains, in the form of a semi-direct product, the
Lorentz group ${\mathcal L} = SO(3,1)$ and the translation group ${\mathcal T}$. The latter acts
transitively on $M$ and its manifold is just $M$. Indeed, Minkowski  spacetime is a homogeneous space
under ${\mathcal P}$, actually the quotient $M \equiv {\mathcal T} = {\mathcal P}/{\mathcal L}$. The
invariance of $M$ under the transformations of ${\mathcal P}$ reflects its uniformity. The Lorentz
subgroup provides an isotropy around a given point of $M$, and translation invariance enforces this
isotropy around any other point.  This is the usual meaning of ``uniformity", in which ${\mathcal T}$
is responsible for the equivalence of all points of spacetime.

Let us then analyze the kinematic group of the de Sitter and anti-de Sitter spacetimes. In the
Cartesian coordinates $\xi^A$, the generators of the infinitesimal de Sitter transformations are
\be
J_{A B} = \eta_{AC} \, \xi^C \, \frac{\partial}{\partial \xi^B} -
\eta_{BC} \, \xi^C \, \frac{\partial}{\partial \xi^A},
\label{dsgene}
\ee
which satisfy the commutation relations
\[
\left[ J_{AB}, J_{CD} \right] = \eta_{BC} J_{AD} + \eta_{AD} J_{BC} - \eta_{BD} J_{AC}
- \eta_{AC} J_{BD}.
%\label{dsal}
\]
In terms of the stereographic coordinates $x^a$, these generators are written as
\be
J_{a b} \equiv L_{ab} =
\eta_{ac} \, x^c \, P_b - \eta_{bc} \, x^c \, P_a
\label{dslore}
\ee
and
\be
J_{4 a} = - {\sf s} \, 
\left(L \, P_a + \frac{{\sf s}}{4 L} \, K_a \right),
\label{dstra}
\ee
where
\be
P_a = \frac{\partial}{\partial x^a} \quad {\rm and} \quad
K_a = \left(2 \eta_{ab} x^b x^c - \sigma^2 \delta_{a}{}^{c}
\right) P_c
\label{cp2} 
\ee
are, respectively, the generators of translations and {\it proper} conformal transformations. For
${\sf s} = -1$, they give rise to the de Sitter group $SO(4,1)$. For ${\sf s} = +1$, they give rise to
the anti-de Sitter group $SO(3,2)$. The generators $J_{a b}$ refers to the Lorentz subgroup
$SO(3,1)$, whereas $J_{a 4}$ define the transitivity on the corresponding homogeneous spaces. Since
$L = [3/(-{\sf s} \, \Lambda)]^{1/2}$ (notice that $-{\sf s} \, \Lambda > 0$ for both de Sitter and
anti-de Sitter cases), we see from Eq.~(\ref{dstra}) that the de Sitter spacetimes are transitive under
a mixture of translations and proper conformal transformations. The relative importance of each one of
these transformations is determined by the value of the cosmological term. In particular, for a
vanishing $\Lambda$, both de Sitter groups are reduced to Poincar\'e, and both de Sitter spaces become
the Minkowski spacetime, which is transitive under ordinary translations. 

\subsection{Other Coordinate Systems}

In the so called {\em global} coordinates $(\tau, \chi, \theta, \phi)$, the de Sitter metric
acquires the form
\be
ds^2 = c^2 d\tau^2 - L^{2} \cosh^2(L^{-1} c \tau) \left[d\chi^2 + \sin^2 \chi
\, (d\theta^2 + \sin^2 \theta d\phi^2)\right].
\label{globalcor}
\ee
In these coordinates, the metric is seen to explicitly depend on the time coordinate. However, there
is a specific coordinate system in which it becomes time independent. This is the so called {\em static}
coordinates $(t, r, \theta, \phi)$, in which the metric has the form
\be
ds^2 = (1- {r^2}/{L^2} ) c^2 dt^2 - \frac{dr^2}{\left(1-r^2/L^2 \right)} -
r^2 (d\theta^2 + \sin^2\theta d\phi^2).
\label{staticor}
\ee
In this form, the de Sitter metric reveals another important property of the de Sitter spacetime,
namely, the existence of a horizon at $r = L$. It also shows that $t$ is a time-like coordinate
only in the region $r< L$. Of course, in the case of a time-depending cosmological term, the 
metric components will also acquire a implicit time dependence.

\subsection{Thermodynamic Properties}

There is a remarkable relation between gravitational horizons and thermodynamic properties. The
classic example is that of the Schwarzschild solution
\be
ds^2 = \left(1- {2 G M}/{c^2 r} \right) c^2 dt^2 - \frac{dr^2}{(1- {2 G M}/{c^2 r})} -
r^2 (d\theta^2 + \sin^2\theta d\phi^2),
\label{swarzsc}
\ee
which represents a black hole of mass $M$. In this case, one can associate with the horizon at $r = 2 G
M/c^2$ a temperature $T_{bh}$ and an entropy $S_{bh}$, given respectively by \cite{bhth}
\be
T_{bh} = \frac{\hbar^2}{8 \pi k_B M l_P^2} \quad {\rm and} \quad
S_{bh} = \frac{k_B A_h}{4 \, l_P^2},
\label{TandS}
\ee
where $k_B$ is the Boltzmann constant, $l_P = \sqrt{G \hbar/ c^3}$ is the Planck length and $A_h$ is the
area of the horizon. Now, as is well known, the phenomenon of black hole evaporation, which consists in
an energy exchange between the black hole and the external space, is a genuine thermodynamical process.
As a consequence, the above quantities are found to be related through the thermodynamic identity
\be
dE_{bh} = T_{bh} \, dS_{bh},
\label{flbh}
\ee
usually called the {\em first law of black hole thermodynamics}. Since in this process the black hole
mass $M$ changes, it can be consistently considered as a variable parameter. In this case, integrating
Eq.~(\ref{flbh}) with $T_{bh}$ and $S_{bh}$ given by Eq.~(\ref{TandS}) yields
\be
E_{bh} = M c^2,
\ee
which is the total energy inside the Schwarzschild horizon.

Due to the common presence of a horizon, in the same way as in the Schwarz\-schild case, it is possible
to attribute thermodynamic features also to the de Sitter horizon \cite{gh}. This result can be
demonstrated mathematically in many different ways, of which the simplest procedure is based on the
relationship between temperature and the Euclidean extension of spacetime. This can be seen by
observing that spacetimes with horizons present a natural analytic continuation from Minkowskian to
Euclidean signature, which is obtained by making $\tau \to i \tau$. If the metric becomes periodic, one
can naturally associate a notion of temperature to such spacetimes. For example, the de Sitter metric
(\ref{globalcor}) can be continued to imaginary time yielding
\be
- ds^2 = c^2 d\tau^2 + L^{2} \cos^2 (L^{-1} c \tau) \left[ d\chi^2 + \sin^2
\chi (d\theta^2 + \sin^2 \theta d\phi^2)\right],
\label{imagds} 
\ee
which is clearly periodic in $c \tau$, with period $2 \pi L$. It then follows that the
temperature
\be
T_{dS} = \frac{\hbar c}{2 \, \pi \, L \, k_B}
\label{dstemp}
\ee
can be associated with the de Sitter spacetime. In a similar fashion, one can also associate to the
de Sitter horizon the entropy
\be
S_{dS} = \frac{k_B \, A_h}{4 \, l_P^2} \equiv  \frac{\pi \, c^3 \, k_B \, L^2}{G \hbar},
\label{dsentro}
\ee
where  $A_h = 4 \pi L^2$ is the area of the horizon.

Although the attribution of temperature and entropy to the de Sitter horizon is quite reasonably
understood, the definition of energy is still highly controversial \cite{paddy}. It is, however, still
possible to make some speculations about this point. First, we note that, for consistency
reasons, whenever a cosmological term is present, an equation of state of the form
\be
p_{dS} = - \varepsilon_{dS} \equiv - \frac{E_{dS}}{V}
\label{steq}
\ee
has to be introduced, where $p_{dS}$ is the pressure, $\varepsilon_{dS}$ is the energy density, and
$V$ is the volume enclosed by the horizon. Accordingly, the first law of the de Sitter thermodynamics
is to be written as
\be
dE_{dS} = T_{dS} \, dS_{dS} - p_{dS} \, dV.
\label{flds}
\ee
For a constant cosmological term, and rendering as implausible the existence of the notion of
evaporation of the de Sitter universe, the energy, entropy and volume will be constant, and the above
equation will be trivially satisfied. However, for a time varying cosmological term \cite{decay}, the
de Sitter parameter $L$ will change with time, and so will do $E_{dS}$, $S_{dS}$, and $V$. Using that
$V \sim L^3$, which implies
\[
\frac{dV}{V} = \frac{3}{L} \, dL,
\]
equation (\ref{flds}) can be integrated to give
\be
E_{dS} = - \frac{c^4 \, L}{2 G}.
\label{dsener}
\ee

Before proceeding further, let us examine the reason for the minus sign in the energy. To begin with, we
observe that there is a fundamental difference between the Schwarzschild and the de Sitter solutions:
whereas the former is valid outside, the latter is valid inside their corresponding horizons. The above
energy, therefore, refers necessarily to the internal side of the de Sitter horizon. Now, if we ascribe
a negative energy to the internal side of the horizon, because the Killing vector $\partial / \partial
t$, used to define energy, changes direction across the horizon, we are forced to ascribe a positive
energy to the external side of the horizon \cite{ssv}. When we do that, a direct comparison can then be
made with the black hole case, whose solution is valid outside the corresponding horizon. Based on this
argument, we can assert that the correct expression for the energy associated to (the external side of)
a de Sitter horizon is
\be
E_{dS} = + \frac{c^4 \, L}{2 G}.
\label{dsener2}
\ee
It is important to remark that only energy changes sign across the horizon. The consistency of this
result can be verified by observing that only for a positive energy the equation of state (\ref{steq})
will give rise to a negative pressure, as required by dark energy.

Using now the analogy with the black hole case, we might expect that the energy associated
to the external side of the horizon coincides with the energy enclosed by the horizon. When there
exists a timelike Killing vector $\xi^\mu = (1, \vec{0})$ associated to time the translation
${\partial}/{\partial t}$, this energy can be written in the form \cite{mottola}
\be
E_{dS} = \int_{r\leq L} \sqrt{h} \; T_{\mu \nu} \; \xi^\mu \; n^\nu \, d^3x,
\label{Ecosmo}
\ee
where $h$ is the determinant of the induced metric on a $t=$ {\it constant} section of the de Sitter
spacetime, $T_{\mu \nu}$ represents the energy-momentum density of the $\Lambda$ term, and $n^\mu =
\xi^\mu/\xi$. Now, in the static coordinates (\ref{staticor}), we have
\[
\xi \equiv |\xi^\mu| = \left(g_{\mu \rho} \, \xi^\mu \xi^\rho \right) =
(g_{00})^{1/2} \quad \mbox{and} \quad \sqrt{h} = (g_{11})^{1/2} \; r^2 \sin \theta.
\]
Making use of the energy-density invariant definition
\[
\varepsilon_{dS} = T_{\mu \nu} \; n^\mu \, n^\nu,
\]
it is then easy to verify that, in order to yield the energy (\ref{dsener2}), we must have
\be
\varepsilon_{dS} = \frac{3 c^4}{8 \pi G L^2}.
\label{enerdensity}
\ee
We see from this expression that the energy density associated with the de Sitter horizon coincides with
the (dark) energy density related to a {\em positive} cosmological term. Notice that, whereas the energy
grows up linearly with an increasing $L$, the energy density falls off with $L^{-2}$. Finally, it is
important to remark that a time-decaying cosmological term implies necessarily that matter be created
inside the horizon \cite{vish}. Of course, as the universe expands and matter is created, spacetime will
no longer be a {\em pure} de Sitter space. However, since the only effect of the presence of matter is
to change the position of the horizon in relation to a pure de Sitter spacetime, it is still possible
to make sense of the above thermodynamic quantities, even in the presence of matter.

%%%%%%%%%%%%%%%%%%%%%%%%%%%%%%%%%%%%%%%%%%%%%%%%%%
\section{The Infinite Cosmological Term Spacetime}

\subsection{Introductory Remarks}

We consider now the In\"on\"u-Wigner contraction \cite{inonu1} limit $\Lambda \to \infty$ (which
corresponds to $L \to 0$) of the de Sitter solutions studied in the previous section.
To begin with, let us remark that, on account of the quotient character of the de Sitter spacetimes,
geometry and algebra are deeply mixed. Any deformation in the algebras and groups, therefore, will
produce concomitant deformations in the imbedded spacetimes. Since the contraction of groups and
algebras is a mathematically well established procedure, we have consequently a rigorous method of
studying limit solutions of Einstein's equation. As an example, and for the sake of completeness, let
us consider first the contraction limit $\Lambda \to 0$ (which corresponds to $L \to \infty$). In this
limit, as is well known, the de Sitter and anti-de Sitter groups are lead to the
Poincar\'e group ${\mathcal P}$, the semi-direct product of the Lorentz ${\mathcal L}$ and the
translation ${\mathcal T}$ groups: ${\mathcal P} = {\mathcal L} \, \oslash \, {\mathcal T}$. Its
generators are $L_{ab}$ and $P_a$, which satisfy the algebra
\ba
\left[L_{ab}, L_{cd}\right] &=& \eta_{bc} \,
L_{ad} + \eta_{ad} \, L_{bc} - \eta_{bd} \,
L_{ac} - \eta_{ac} \, L_{bd}, \label{ll} \nonumber \\
\left[ P_{a}, L_{cd}\right]&=&\eta_{ac}  P_{d} -
\eta_{ad}  P_{c}, \label{pl} \nonumber \\
\left[ P_{a}, P_{c}\right]&=&0. \nonumber
\label{pp}
\ea
As a result of this group deformation, the de Sitter metric (\ref{44}) is easily seen to become the
Minkowski metric, and the Riemann, Ricci and scalar curvature tensors are found to vanish. This means
essentially that one obtains the flat Minkowski space $M = {\mathcal P}/{\mathcal L}$, which is {\em
transitive} under ordinary translations.

\subsection{Infinite Cosmological--Term Limit}

In the contraction limit  $\Lambda \to \infty$, the de Sitter groups are lead to the so called
{\it second} or {\it conformal} Poincar\'e group ${\mathcal Q}$, the semi-direct product between
Lorentz ${\mathcal L}$ and the proper conformal group ${\mathcal C}$, that is, ${\mathcal Q} =
{\mathcal L} \, \oslash \, {\mathcal C}$ \cite{ap1}. Its generators are $L_{ab}$ and $K_a$, which
satisfy the same commutation relations of the Poincar\'e Lie algebra:
\ba
\left[L_{ab}, L_{cd}\right] &=& \eta_{bc} \,
L_{ad} + \eta_{ad} \, L_{bc} - \eta_{bd} \,
L_{ac} - \eta_{ac} \, L_{bd}, \label{llbis} \nonumber \\
\left[ K_{a}, L_{cd}\right]&=&\eta_{ac}  K_{d} -
\eta_{ad}  K_{c}, \label{kl} \nonumber \\
\left[ K_{a}, K_{c}\right]&=&0. \nonumber
\label{kk}
\ea
It should be mentioned, however, that the Lie group corresponding to this algebra is completely
different from the ordinary Poincar\'e group ${\mathcal P}$.

\begin{figure}[h]
\begin{center}
\includegraphics[height=6cm,width=12cm]{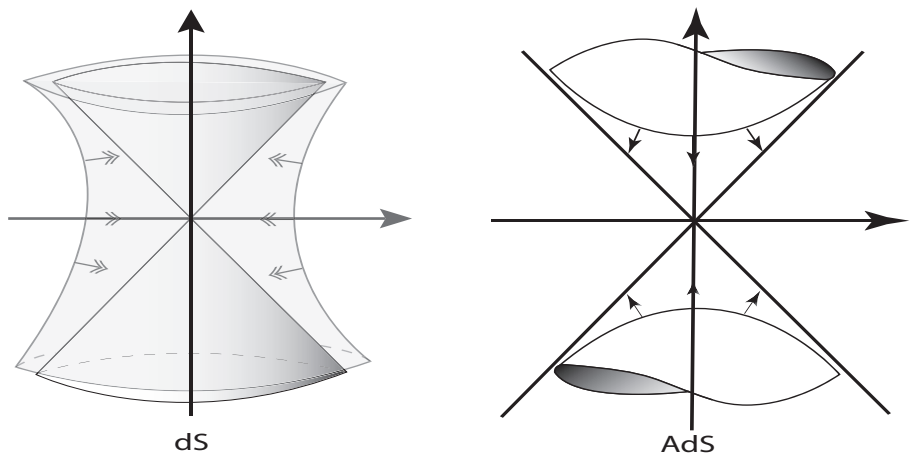}
\end{center}
\caption{The figure on the left represents the de Sitter space contraction towards the cone-space. The
figure on the right represents the anti-de Sitter space contraction towards the same cone-space. Notice
that the de Sitter space approaches the cone from outside, whereas the anti-de Sitter approaches the
cone from inside (for easy visualization, two dimensions have been suppressed in these figures).}
\label{f1}
\end{figure}

As already remarked, the above group deformation will produce concomitant changes in the imbedded
spacetimes. In fact, we see from Eq.~(\ref{dspace2}) that in the limit of an infinite cosmological term
($L \to 0$), the de Sitter and the anti-de Sitter spaces are both led to a four-dimensional cone-space,
which we denote by $N$. This process is pictorially shown in Figure~\ref{f1}, from where we see that,
whereas the de Sitter space approaches the cone from outside, the anti-de Sitter approaches the cone
from inside.

%%%%%%%%%%%%%%%%%
%%%%%%%%%%%%%%%%%%

Now, it can be seen that, in the limit $L \to 0$, the de Sitter metric becomes singular everywhere in
$N$:
\be
\lim_{L\to 0} \, g_{\mu \nu} = 0, \qquad
\lim_{L\to 0} \, g^{\mu \nu} \to \infty.
\label{singu}
\ee
As a consequence, no {\em ordinary} invariant interval can be defined on $N$, which means that
neither the usual notions of space distance and time interval hold. Nevertheless, the Levi-Civita
connection (\ref{46}) is well defined, and consequently the Riemann curvature tensor is also well
defined. As an explicit computation shows, whereas both the Riemann and the Ricci curvature tensors
vanish for $L \to 0$, the scalar curvature becomes infinity:
\be
\lim_{L \to 0} \, R \to \infty.
\ee
Vanishing Riemann and Ricci tensors with an infinity scalar curvature is a characteristic property of a
spacetime with an infinite cosmological term \cite{congal}.

\subsection{Kinematic Group: Transitivity}

Like the Minkowski spacetime $M$, the cone-space $N$ is a homogeneous space, but under ${\mathcal Q}$:
$N = {\mathcal Q}/{\mathcal L}$. The kinematical group ${\mathcal Q}$, as the Poincar\'e group, has
the Lorentz group ${\mathcal L}$ as the subgroup accounting for the isotropy of $N$. However, the
proper conformal transformations introduce a new kind of homogeneity: instead of the ordinary
translations  defining homogeneity on Minkowski spacetime, all points of $N$ are equivalent through
proper conformal transformations. In other words, the cone-space $N$ is transitive under proper
conformal transformations. The Poincar\'e ${\mathcal P}$ and the conformal Poincar\'e ${\mathcal Q}$
groups can then be considered as dual to each other in the sense that they are the isometry groups of
spacetimes characterized respectively by a vanishing and an infinite cosmological term. 

A crucial point is to notice that there are two notions of space distance and time interval involved:
the usual one, related to translations and defining the transitivity of Minkowski spacetime, and the
conformal one, related to the transitivity of spacetimes with an infinite cosmological term. In the
intermediate case of a finite cosmological term, the transitivity of the corresponding de Sitter
spacetime is given by a mixture of these two notions, as can be seen in the de Sitter ``translation''
generators (\ref{dstra}). The relative importance between the translation and the proper conformal
generators is determined by the value of the cosmological term. For $\Lambda \to 0$, the de Sitter
spacetime becomes the Minkowski space $M$, which is transitive under ordinary spacetime translations.
For $\Lambda \to \infty$, the de Sitter spacetime becomes the four-dimensional cone-space $N$, which
is transitive under proper conformal transformations. Two generic points on $N$ cannot be taken one
into the other by ordinary translations, but can be related by proper conformal transformations. On
account of this conformal transitivity, the cone-space $N$ can be said to be {\em conformally infinite}.

Despite presenting all properties of an ordinary spacetime singularity, the vertex is smoothly
connected with all other points of the cone-space through proper conformal transformations (the
conformal infiniteness alluded to a\-bove). Although the ordinary notions of space distance and time
interval fail on $N$, the corresponding notions of conformal space and  conformal time can be defined
\cite{aap}. This means essentially that, in addition to the ordinary metric $g_{\mu \nu}$ (which
becomes singular in the contraction limit $\Lambda \to \infty$), it is possible to define another
metric on $N$, which will be responsible for the conformal notions of space distance and time interval.

\subsection{Conformal Invariant Metric}

Defining the reciprocal ambient space coordinate \cite{foot2}
\be
\chi^a = \xi^a / 4 L^2,
\ee
it is easy to see that, in the contraction limit $L \to 0$, the stereographic projection (\ref{xix})
becomes the spacetime inversion, which in the case of a positive cosmological term (${\sf s}=-1$)
is given by
\be
\chi^a = - \frac{x^a}{\sigma^2}.
\ee
Now, this transformation is well known to relate translations with proper conformal transformations
\cite{coleman}. In fact, under the spacetime inversion
\be
x^a \to \bar{x}^a = - \frac{x^a}{\sigma^2},
\label{inversion}
\ee
the translation generators transform according to
\be
P_a \rightarrow \bar{P}_a = K_a.
\ee
Since the Lorentz generators are invariant,
\begin{equation}
L_{ab} \rightarrow \bar{L}_{ab} = L_{ab},
\end{equation}
the inversion transformation is seen to relate actually the Poincar\'e with the conformal Poincar\'e
groups. Now, under this transformation, the Minkowski space $M$ is transformed into the
four-dimensional cone-space $N$, and vice-versa. Notice, for example, that the points at infinity of
$M$ are lead to the vertex of the cone-space $N$, and those on the light-cone of $M$ become the
infinity of $N$. Minkowski and the cone spacetimes can, therefore, be considered as {\it dual} to each
other in the sense that their geometries are determined, respectively, by a vanishing and an infinite
cosmological term. The {\em duality} transformation connecting these two geometries is the spacetime
inversion (\ref{inversion}).

On account of the above described relation, by applying the duality transformation (\ref{inversion})
to the Minkowski metric $\eta^{ab}$, it is possible to obtain a conformal invariant metric
${\bar{\eta}}_{ab}$ on the cone-space $N$, which is found to be
\be
{\bar{\eta}}_{ab} = {\sigma}^{-4} \, \eta_{ab}; \qquad {\bar{\eta}}^{ab} =
{\sigma}^{4} \, \eta^{ab}.
\label{Nmetric}
\ee
Therefore, if
\be
ds^2 = \eta_{ab} \, dx^a dx^b
\ee
is the Minkowski interval, the corresponding cone-space ``conformal interval'' will be
\be
d\bar{s}^2 = {\eta}_{ab} \, d\bar{x}^a d\bar{x}^b = {\bar{\eta}}_{ab} \, d{x}^a d{x}^b.
\label{confin}
\ee
As a simple calculation shows, the metric $\bar{\eta}_{ab}$ is in fact invariant under the conformal
Poincar\'e group ${\mathcal Q}$.

\subsection{Thermodynamic Properties}

Let us now analyze the behavior of the thermodynamic quantities in the limit of an infinite
cosmological term ($L \rightarrow 0$). In that limit the temperature, according to definition
(\ref{dstemp}), becomes infinity: $T_{dS} \rightarrow \infty$. The entropy, on the other hand,
according to the expression (\ref{dsentro}), vanishes: $S_{dS} = 0$. And finally, on account of the
definition (\ref{dsener}), the energy associated with the de Sitter horizon vanishes identically:
$E_{dS} = 0$. It is important to observe that, although the energy vanishes, the energy density,
according to the expression (\ref{enerdensity}), becomes infinity: $\varepsilon_{dS} \to \infty$. The
reason for this behavior is that the volume delimited by the horizon vanishes faster than the energy.
We can thus say that the above initial conditions are consistent with what would be expected for a
big-bang universe. It is also important to remark that, as the Minkowski spacetime is obtained in the
limit of a vanishing cosmological term ($L \rightarrow \infty$), we can associate to it a vanishing
temperature, an infinite entropy \cite{foot3} and an infinite energy, with the energy density going to
zero. Furthermore, the horizon associated to a Minkowski spacetime becomes infinite, which is the same
as no horizon at all.

%%%%%%%%%%%%%%%%%%%%%%%
\section{Final Remarks}

The spacetime associated with an infinite cosmological term has a quite peculiar geometry. To
understand it, let us consider the Minkowski spacetime. As is well known, it is transitive under
spacetime translations. In the presence of a positive cosmological term, Minkowski is transformed into
a de Sitter spacetime, which is transitive under a mixture of translation and proper conformal
transformations. In a de Sitter spacetime, therefore, there are two notions of space distance and time
interval: the usual one, related to translations, and the conformal one, related to the proper
conformal generators. The relative importance of these two notions, as can be seen from the de Sitter
generators (\ref{dstra}), is determined by the value of $\Lambda$. For $\Lambda \rightarrow \infty$,
the de Sitter spacetime becomes the four-dimensional singular cone-space $N$, transitive under proper
conformal transformations only. Its kinematical group of motion is the {\em second}, or {\em conformal}
Poincar\'e group, the semi-direct product of Lorentz and proper conformal groups. In this spacetime,
therefore, the usual concepts of space distance and time interval break down, a property also revealed
by the fact that the usual metric tensor is singular everywhere on $N$. Another way to see this
property is to observe that the causal domain of any single inertial observer is bounded by its
horizon, whose radius goes to zero for an infinite cosmological constant. In other words, its causal
domain collapses to a point. Notice that, although the collapse is observer-dependent, it is meaningful
since any observer will experience it.

In spite of the lack of the usual notions of space distance and time interval, it is possible to define
a conformal invariant metric on the cone space $N$, which defines a conformal interval. This means that
any two points will not be connected by the usual concepts of space and time intervals, but by a {\em
conformal} notion of space and time intervals. Except for the singular vertex, therefore, the cone space
$N$ can be said to be conformally smooth and infinite. It is a new example of a maximally-symmetric
spacetime, with a well defined group of motions---the conformal Poincar\'e group.

A possible application of these results refers to the cosmology of the early Universe. According to the
standard big-bang model, the universe started from a singular point. As general relativity breaks down
at any singularity, the initial condition of the universe cannot be characterized as a particular
solution of the theory governing the universe dynamics. It is consequently left to be chosen freely,
and often considered to be beyond the scope of the known physical theories. On the other hand, recent
observations \cite{recent} indicate the presence of a nonvanishing cosmological term $\Lambda$ in our
Universe. Furthermore, inflation requires a very high value for $\Lambda$ at the early stages of the
universe \cite{amm}. It is consequently appealing to assume an infinite primordial $\Lambda$ at some
initial cosmological time \cite{mwa}, followed by a decaying but still large cosmological term, which
would drive inflation. This means to assume the initial condition of the universe as the spacetime
defined by the de Sitter solution in the limit of an infinite $\Lambda$. The initial state of the
universe would, in this way, become linked to the limit of a specific solution of Einstein's equation,
and in consequence consistently established as part of the theory \cite{bojowald}. Accordingly, all
energy of the universe would be initially in the form of dark energy. It is important to notice that
this spacetime presents also thermodynamic properties that fit quite reasonably with what one would
expect for a consistent initial state for the universe. In fact, it has an infinite temperature, and
vanishing entropy and energy. The energy density, however, is infinite. We can then say that such a
singular conformal spacetime presents the basic properties (thermodynamic and geometric) to be
considered as a consistent initial state for a big-bang universe, in which all energy is dark.

From the above considerations, the {\em birth} of the usual concepts of space and time can be related
with the transition from an infinite to a finite cosmological term. In fact, suppose for example that
quantum fluctuations near the cone-space vertex singularity cause the de Sitter length-parameter $L$ to
assume, instead of zero, the Planck length: $L = l_P \simeq 1.6 \times 10^{-33}\;{\rm cm}$. When this
occurs, the singularity disappears, and the cone-space is transformed into a de Sitter (anti-de Sitter)
spacetime with a very high positive (negative) cosmological term. For a quantum fluctuation generating
a negative cosmological term, the resulting anti-de Sitter spacetime will be pushed back to the
original singular state by the highly attractive force produced by the negative $\Lambda$. This means
that the cone-space $N$ might be stable under quantum fluctuations generating a negative $\Lambda$.
However, for a quantum fluctuation generating a positive cosmological term, the resulting de Sitter
spacetime will be unstable due to the repulsive force produced by $\Lambda$, and  can eventually give
rise to a rapidly expanding universe. In that plausible case, the ordinary spacetime translations will
immediately show up, together with proper conformal transformations, in the de Sitter {\em translation}
generators (\ref{dstra}). At this moment, our usual notions of space distance and time interval emerge,
though the corresponding conformal notions still remain as the most important part of the de Sitter
{\em translation} generators. The universe begins to expand through ordinary space, and ordinary time
begins to flow. Of course, the model presupposes a time-decaying cosmological term, which in turn
implies necessarily that matter be created inside the de Sitter horizon \cite{vish}. This means that
not only ordinary space and time, but also the whole matter content of the universe would arise from
the primordial infinite cosmological term \cite{relax}. 

%%%%%%%%%%%%%%%%%%%%%%%
\begin{acknowledgments}
The authors would like to thank FAPESP-Brazil, CAPES-Brazil, and CNPq-Brazil for financial support.
\end{acknowledgments}


\begin{thebibliography}{99}
\bibitem{foot2}
The factor 4 has been introduced to compensate the same factor appearing in the de Sitter generators
(\ref{dstra}).
\bibitem{foot3}
The concept of entropy is intimately related to the notion of information. In this sense, as no single
observer in de Sitter can access the entire spacetime, she/he has limited information, which appears in
the form of a finite value for the corresponding entropy. This is qualitatively different from
Minkowski space, where a timelike observer will eventually have the entire history of the universe in
the past light-cone. This corresponds to an infinite information, and consequently to an infinite
entropy.
\end{thebibliography}
\end{document}